\documentclass{ws-ijmpd}

\begin{document}

\markboth{F. Tavecchio}
{Gamma-ray emission from AGNs}

%
\catchline{}{}{}{}{}
%

\title{GAMMA-RAY EMISSION FROM AGNS (SPECIAL FOCUS ON BL LAC OBJECTS)}

\author{F. TAVECCHIO}

\address{INAF-OAB, Via Bianchi 46\\ 
I-23807, Merate, LC, Italy\\
fabrizio.tavecchio@brera.inaf.it}



\maketitle

\begin{history}
\received{Day Month Year}
\revised{Day Month Year}
\comby{Managing Editor}
\end{history}

\begin{abstract}
Blazars, radio-loud active galactic nuclei with the relativistic jet closely aligned with the line of sight, dominate the extragalactic sky observed at gamma-ray energies, above 100 MeV. We discuss some of the emission properties of these sources, focusing in particular on the ``blazar sequence'' and the interpretative models of the high energy emission of BL Lac objects.
\end{abstract}


\section{Introduction}	

The extragalactic gamma-ray sky at energies above 100 MeV is mainly populated by blazars, radio loud active galactic nuclei (AGNs) whose emission is dominated by the relativistically boosted non-thermal continuum produced in a relativistic jet  (bulk Lorentz factor $\Gamma=10-20$) pointing (typical angles less than 5 degrees) toward the Earth. Blazars are among the most powerful and violently variable sources in the Universe: their emission covers all the electromagnetic spectrum, from the radio to the gamma-ray band, and their bolometric output (often concentrated in the gamma-ray band) can exceeds $L>10^{48}$ erg s$^{-1}$. The variability timescale can be as short as few minutes\cite{albert07}\cdash\cite{aharonian07}. The interest for these sources is driven by the possibility to probe the innermost regions and the surroundings of the relativistic jets, close to the acceleration and collimation region ($d\sim 10^2-10^3$ gravitational radii). 

In the past decade  EGRET onboard {\it CGRO} detected almost 70 blazars. Of the 106 sources in the recently released list\cite{abdo09} (LBAS) of the AGNs detected with high significance by the Large Area Telescope (LAT) onboard the  {\it Fermi Gamma Ray Telescope}\cite{fermi} in the first three months of operation (August-October 2008), 104 are blazars, the remaining two sources being the radiogalaxies Perseus A and Centaurus A.  Note that one of the sources classified as a blazar is a radio loud Narrow Line Seyfert 1 galaxy, the first object of this class revealed at gamma rays\cite{nls1}\cdash\cite{foschini}.
The gamma-ray emission can extend up to the VHE band ($E>100$ GeV), at which the observations are performed by the Cherenkov telescopes on the ground. Also in this band blazars dominate the extragalactic sky. The VHE emission can be used to derive limits on the density of the IR-optical backgrounds exploiting the modification of the spectrum due to the absorption through the pair production process\cite{stecker}\cdash\cite{mazin}.

Due to the limited space, in the following I will focus the attention on some specific issues: i) the blazar sequence and its physical meaning; ii) the high-energy view of the the BL Lac objects and some problems possibly requiring a revision of the simple one-zone emission models.

\section{SEDs and the blazar sequence}

\begin{figure}[pb]
\centerline{
\psfig{file=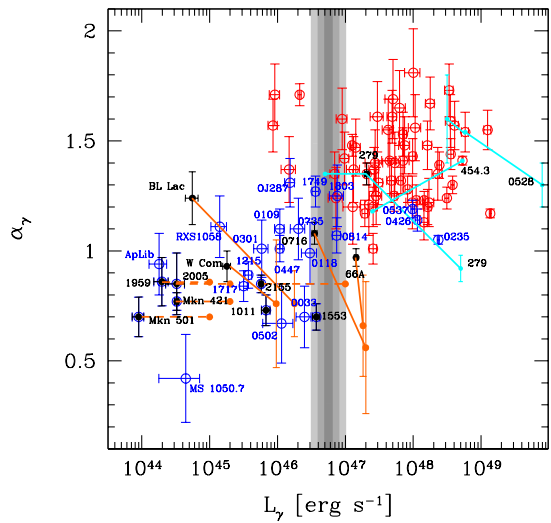,width=8cm}
}

\caption{Energy index of the gamma-ray spectrum, $\alpha_{\gamma}$, versus the gamma-ray luminosity for all the blazars detected with high significance by the LAT instrument onboard {\it Fermi} in its three months of observations. Open circles represent BL Lac objects, open squares flat spectrum radio quasars, filled symbols sources detected in the TeV band. Solid lines connect points corresponding to different emission states of the same source. A trend between the two quantities is clearly visible, together with a sharp division between the two classes of objects occurring at  a luminosity $L_{\gamma}\sim 10^{47}$ erg s$^{-1}$. These properties confirms the expectation of the so called ``blazar sequence'' (see text for details). 
   \label{f1}}
\end{figure}

The spectral energy distribution (SED) of blazars displays two broad peaks, interpreted as due to synchrotron and inverse Compton (IC) mechanisms in the popular leptonic models\cite{mgc92}\cdash\cite{sikora} (see, e.g., Ref.~\refcite{muecke} for alternatives hadronic models). As shown almost ten years ago by Fossati et al.\cite{fossati}, averaged SEDs of  blazars in different bins of radio luminosity show a trend with the radio (or the bolometric) luminosity: specifically, flat spectrum radio quasars (FSRQs), the most powerful blazars, showing intense broad emission lines and sometimes a thermal bump in their optical spectra, have both peaks located at low frequencies, submm-IR and MeV for the synchrotron and IC peak, respectively. On the contrary, in the less powerful BL Lac objects (characterized by very weak emission lines), the two bumps peak at UV-X-ray frequencies and above 100 GeV. The existence and the validity of such a trend (dubbed ``blazar sequence'') has been hotly debated in the past\cite{padovanirev}. A possible {\it caveat} concerns the use of the EGRET data to derive the averaged gamma-ray spectra, since the limited sensitivity likely introduces a bias toward flaring states. One argument of the critics is the claimed existence of ``blue quasars'', FSRQs with peaks at high frequencies, violating the sequence. However, so far none of the blue quasars candidates has been proven to be a robust case\cite{landt}\cdash\cite{maraschi08}. 

The three-month LAT data confirm with a larger statistics\cite{gmt09} one of the basic expectations of the blazar sequence, namely the fact that FSRQs should have (on average) soft GeV spectra (with energy index $\alpha_{\gamma}>1$) whereas BL Lac should show hard spectra ($\alpha_{\gamma}<1$).  As shown in Fig.\ref{f1} this prediction is confirmed by the new data, showing not only the expected dicotomy in the spectral slope, but also a well definite trend between the gamma-ray luminosity and the slope of the gamma-ray spectrum. The small ``thickness'' (or even the very existence) of the correlation is even surprising, considering the huge variability possible for both quantities in the same source. This is  shown for some selected objects by the  solid lines line connecting different emission states of the same source. In Ref.~\refcite{gmt09} we proposed that the divide in the gamma-ray luminosity (a good proxy for the bolometric output of the jet) can be associated to a transition in the properties of the accretion flow on the central BH, transiting from a standard Shakura-Sunyaev disk at high luminosities, to an inefficient accretion regime (ADAF\cite{adaf}, ADIOS\cite{adios}). The transition occurs at the level of an accretion rate around $\dot{m}\sim 10^{-2}$ in Eddington units, in agreement with the expectation from the theory\cite{mahadevan} and the analogous scenario proposed to account for the FRI/FRII division among radiogalaxies\cite{ggfr1fr2}.

The existence and the characteristics of the blazar sequence call for a robust and simple explanation. Interestingly, the spectral sequence is associated to a (even more robust) sequence in some of the fundamental quantities of the emission region, as modeled with one-zone leptonic models\cite{ghi08}. In particular, the energy of the electrons emitting at the peaks of the SEDs shows a strong correlation with the energy density (radiative plus magnetic) in the source. The most direct interpretation is that in sources with a large energy density the severe radiative cooling prevents the electrons to reach high energy, determining a low energy frequency of the peaks in the SEDs. In sources in which electrons are characterized small radiative losses the acceleration mechanism can instead bring electrons to higher energies, determining large peak frequencies. Electrons in the jet of FSRQs are commonly assumed to have a rather high cooling rate, since the environment external to the jet is expected to be rather rich of soft photons, determining severe radiative losses through Compton emission. BL Lac objects, instead, are characterized by an environment poor of radiation (and also possibly by a radiatively inefficient accretion flow) and thus electrons can be energized to very high energy, the cooling being dominated by synchrotron and synchrotron-self Compton losses. We refer to Ref.~\refcite{celotti} for a deeper discussion.

\section{BL Lac objects at high energy: GeV and TeV emission}

As discussed above, jets in BL Lac objects are expected to propagate in a ``clean'' environment and the emission we observed from them is likely due to synchrotron and SSC mechanisms. Being less prone to depend on the details specifying the location of the emission zone (e.g. L. Stawarz and L. Costamante, these proceedings) and the conditions of the surrounding environment, the emission from BL Lac objects seems to be the ideal tool to investigate the physics of the particle acceleration and cooling processes acting in jets. Moreover, BL Lac objects are interesting sources since the high energy component the SED is located at the crossroad between the band covered by LAT ($E<100$ GeV) and by ground based Cherenkov telescopes ($E>50$ GeV). It is thus possible to probe the entire high-energy emission, fully covering the bump before and after the IC peak (e.g. Ref.\refcite{2155}).

\begin{figure}[pb]
\vspace*{-3cm}
\centerline{\psfig{file=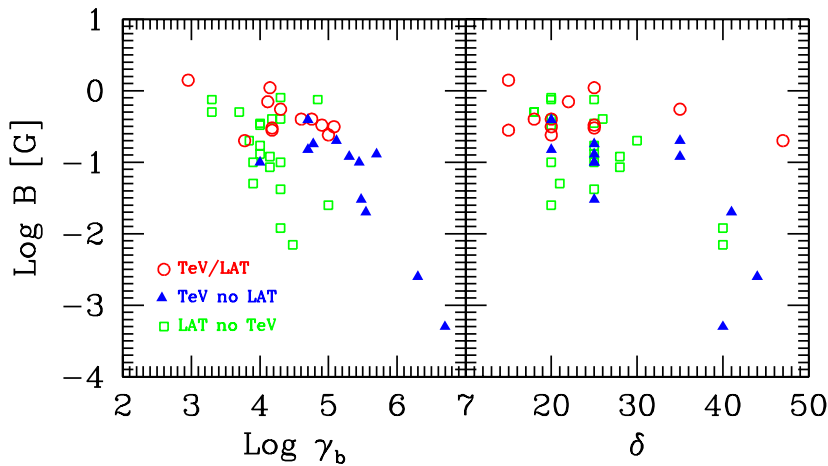,width=13cm}}
\vspace*{-3cm}
\caption{Magnetic field versus the Lorentz factor of the electrons emitting at the SED peak (left) and the Doppler factor of the emitting region (right) derived for the BL Lacs using the one-zone SSC model. While most of the sources are characterized by magnetic fields around 0.1-1 G, Lorentz factors in the range $\gamma_{\rm b}=10^3-10^4$ and $15<\delta<30$, there are some notable exceptions, requiring larger $\delta$ and $\gamma _b$ and smaller magnetic fields.  Different symbols indicate  the instrument (LAT or TeV telescopes) the detected the source. \label{f2}}
\end{figure}

To investigate the global properties of the BL Lac objects revealed at high and very high energies we considered\cite{bllacs} all the BL Lac objects in the LBAS list, including also 12 sources detected at TeV energies. To this sample we add the remaining 12 BL Lac detected at TeV energies (before the end of  2009 September) not detected by LAT.

For all these sources we assembled the SED using the high energy data, historical data and, when available the (often nearly simultaneous) optical-UV and X-ray data from the {\it Swift} satellite. We reproduced the SED with a simple one-zone synchrotron-SSC model, deriving some of the basic physical quantities. For simplicity, the model assumes a spherical homogeneous region and the relativistic electrons are assumed to follow a broken power law distribution in energy with the break at the Lorentz factor $\gamma _{\rm b}$.
In Fig.\ref{f2} we report the values of the magnetic field $B$ versus $\gamma _{\rm b}$ (corresponding to the energy of the electrons emitting most of the power, close to the SED peaks) and the Doppler relativistic factor, $\delta$. Most of the sources occupy a region specified by a magnetic field in the range $B=0.1-1$ G, $\gamma _{\rm b}=10^3-10^5$, $\delta=15-30$. However, there are some notable exceptions, requiring very small magnetic fields ($B\sim 10^{-2}$) G, large $\gamma _{\rm b}$ ($10^6-10^7$) and/or $\delta>40$. 

At a closer look these ``outliers'' can be divided into two classes: i) sources showing a moderately hard LAT spectrum with the synchrotron peak in the optical-UV band; ii) TeV BL Lacs not detected by LAT characterized by extremely hard (de-absorbed) TeV spectra.

\subsection{Hard GeV spectra and low frequency synchrotron peaks}

\begin{figure}[tb]
\vspace*{-1cm}
\centerline{\psfig{file=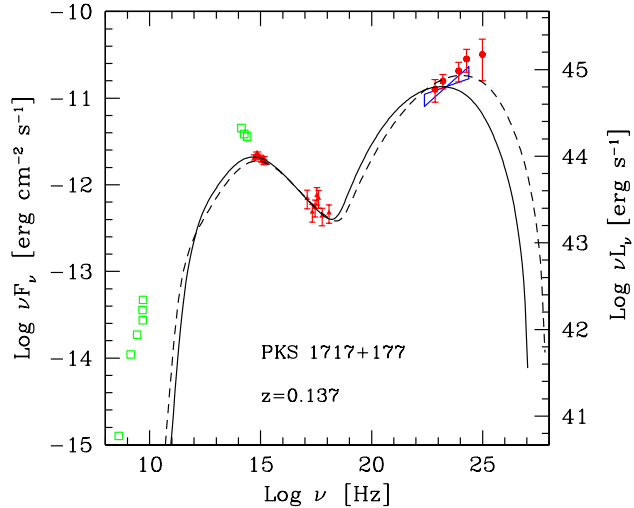,width=9cm}}
\vspace*{-1cm}
\caption{SED of  PKS 1717+177. Optical-X-ray data are from {\it Swift} (January 2009). The ``butterfly'' reports the power-law fit of the first 3 month of LAT data. The corresponding data-points, have been obtained using the the standard analysis procedure for LAT data. The solid line is the result of the one-zone synchrotron-SSC model with a Doppler factor $\delta=20$, the dashed line, instead, has been calculated assuming a somewhat extreme value of $\delta=50$. \label{f3}}
\end{figure}

The SED of one of the BL Lacs belonging to the group i), PKS 1717+177, is given in Fig.\ref{f3} (adapted from Ref.\refcite{bllacs}). Besides the LAT ``butterfly'', calculated with the spectral data of the LBAS catalogue\cite{abdo09}, we report the datapoints derived by analyzing the LAT data with the standard procedure\footnote{{\tt http://fermi.gsfc.nasa.gov/ssc/data/analysis/}}. The points confirm that the spectrum continues to be hard up to the highest energies.

For these kind of sources the large separation between the synchrotron (optical-UV) and the SSC ($E>50$ GeV) peaks in the SED inevitably  requires large Doppler factors\cite{m87}.  In the case of  PKS 1717+177 we show the model resulting by using  a moderate value of the Doppler factor,  $\delta=20$ (solid line): the fit of the LAT data is clearly poor. Using a somewhat extreme value of $\delta=50$ (dashed line) clearly we obtain a better agreement with the LAT data, although the LAT spectrum would be better reproduced with a peak beyond 50 GeV. 
With a word of caution due to the non simultaneity of the {\it Swift} and LAT data, we conclude that the simplest version of the SSC model would clearly be in trouble with these sources as far as we assume that $\delta$ cannot exceed 50.


A viable solution to alleviate the problem represented by the large Doppler factors is offered by the model assuming that the emission zone is not a single homogeneous region, but jet is structured, with a fast ($\Gamma=15-20$) core (or {\it spine}) surrounded by a slower ($\Gamma=3-5$) layer.
These models were originally introduced to account for the discrepancy between the large Doppler factors inferred in some of the TeV BL Lacs and the slow speeds measured in their jets at VLBI scale\cite{ghi05} and have been subsequently adopted to model the high energy emission of radiogalaxies\cite{m87}.
The key ingredient is the radiative coupling between the spine and the layer: both components see the emission from the other relativistically boosted (because of the relative speed). The IC emission is therefore amplified with respect to the SSC case. In particular, if the emission from the layer contains more soft photons than that of the spine, the corresponding IC-scattered photons can reach higher energy than the SSC photons, whose energy is strongly limited by the rapid decline of the Klein-Nishina cross-section. The emission from the spine could then show a high-energy tail from this kind of external Compton component, accounting for the observed LAT spectrum. A full discussion, including more detailed  models is in preparation.

\subsection{Hard TeV spectra}

\begin{figure}[tb]
\vspace*{-1.5cm}
\centerline{\psfig{file=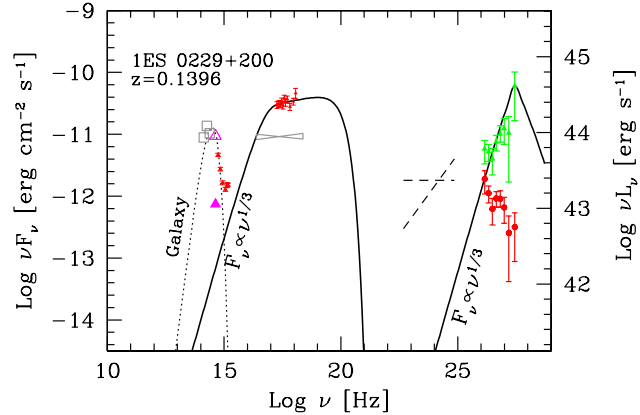,width=9.cm}}
\vspace*{-1.5cm}
\caption{SED of the BL Lac 1ES 0229+200, assembled with historical, {\it Swift} and HESS data. The optical-UV data track the emission from the host galaxy (dotted line). The solid line shows the result of the calculation with a synchrotron-SSC model assuming a simple truncated power law form for the electron energy distribution, with a large value for the minimum Lorentz factor, $\gamma _{\rm min}=8.5\times 10^5$. The dashed lines report the LAT upper limits assuming a photon index of $\Gamma=2$ or 1.5. \label{f4}}
\end{figure}

Among TeV BL Lacs there are a handful of peculiar sources showing very hard TeV spectra (when the observed spectrum is properly deabsorbed to take into account the absoprtion by photons of the cosmic IR-optical background). The SED of one of the most extreme cases, 1ES 0229+200, is reported in Fig.\ref{f4}\cite{0229}. 

In the framework of the one-zone SSC model such hard spectra are difficult to obtain, since the decreasing of the scattering efficiency in the Klein-Nishina regime leads to soft spectra. However, extremely hard spectra can be obtained\cite{katar} if the emitting relativistic electrons have a large value of the minimum electron energy $\gamma_{\rm min}mc^2$: in this case the synchrotron emission, below the typical frequency of the low-energy electrons, $\nu _{\rm s,min}\simeq 3\times 10^6 B \gamma _{\rm min}^2 \delta $ Hz, will display the typical $F(\nu) \propto \nu^{1/3}$ spectrum. The {\it same hard slope} will be reflected by the SSC emission below the peak. 


In Fig.\ref{f4} we report the optical-UV and X-ray data obtained by {\it Swift}, showing that a hard spectrum below the soft X-ray band is required by the data, supporting the possibility discussed above. Clearly, another strong prediction of this scenario is that the level of the emission in the GeV band should be rather low. Consistently, all the sources featuring such hard TeV spectra have not been detected in the first months of Fermi. In a recent paper\cite{abdobllacs}, the LAT collaboration reports a  deep (5.5 months of observations) upper limit for 1ES 0229+200 (that we plot in Fig.\ref{f4}, dashed lines), fully consistent with the theoretical expectations. If our scenario is correct, these sources would be the most efficient cosmic accelerator known, able to push virtually all the electrons at very high energy.

\section{Concluding remarks}

\noindent
1) The ``blazar sequence'' is confirmed and even strengthened by the data from LAT. I would like to remark, however, that  deeper data will start to probe a regime of luminosities populated be a mix of sources. In particular, blazars with small black holes or with jets slightly misaligned will introduce some confusion, making the discussion more difficult.

\noindent
2) The one zone synchrotron-SSC model for BL Lacs gives satisfactorily result for most of the sources when we use almost simultaneous data. However, for a real test of the model it is important to obtain really simultaneous data.

\noindent
3) The BL Lac mainly populating the LAT sky are, obviously, those with a prominent GeV emission. However, there is an interesting population of BL Lacs, characterized by very hard GeV-TeV spectra that are not easily detected by LAT. Any discussion on the high-energy emission of BL Lacs  based only on LAT would exclude these sources that could be the most efficient machines accelerating the relativistic electrons.

\section*{Acknowledgments}

I am indebted with G. Ghisellini, L. Maraschi, G. Ghirlanda and L. Foschini for years of fruitful collaboration. I thank the organizers for the very fruitful and pleasant meeting. This work was partly financially supported by a 2007 COFIN-MIUR grant and by ASI under contract ASI I/088/06/.




\begin{thebibliography}{0}    

\bibitem{abdo09}  A. A. Abdo et al. {\it ApJ} {\bf 700} (2009) 597.

\bibitem{fermi}  W. B. Atwood et al. {\it ApJ} {\bf 697} (2009) 1071.

\bibitem{nls1}  A. A. Abdo et al. {\it ApJ} {\bf 699} (2009) 976.

\bibitem{foschini}  L. Foschini et al., to appear in {\it Proc. Accretion and Ejection in AGNs: a global view}, eds. 
L. Maraschi, G. Ghisellini, R. Della Ceca, F. Tavecchio (ASP Conf. series) (arXiv:0908.3313).

\bibitem{albert07}  J. Albert et al. {\it ApJ} {\bf 669} (2007) 862.

\bibitem{aharonian07}  F. Aharonian et al. {\it ApJ Lett.} {\bf 664} (2007) L71.

\bibitem{stecker}  F. W. Stecker et al. {\it ApJ Lett.} {\bf 390} (1992) L49.

\bibitem{mazin}  D. Mazin, M. Raue {\it A\&A} {\bf 471} (2007) 439.

\bibitem{mgc92}  L. Maraschi, G. Ghisellini, A. Celotti {\it ApJ Lett.} {\bf 397} (1992) L5.

\bibitem{sikora}  M. Sikora et al. {\it ApJ}  {\bf 421} (1994) 153.

\bibitem{muecke}  A. Muecke et al. {\it Astropart. Phys.} {\bf 18} (2003) 593.

\bibitem{fossati}  G. Fossati et al. {\it MNRAS} {\bf 299} (1998) 433.

\bibitem{padovanirev}  G. Ghisellini, F. Tavecchio {\it MNRAS} {\bf 387} (2008) 1669.

\bibitem{landt} H. Landt et al. {\it ApJ} {\bf 676} (2008) 87.

\bibitem{maraschi08} L. Maraschi et al. {\it MNRAS} {\bf 391} (2008) 1981.

\bibitem{gmt09}  G. Ghisellini, L. Maraschi, F. Tavecchio {\it MNRAS} {\bf 396} (2009) L105.

\bibitem{adaf} R. Narayan et al. {\it ApJ} {\bf 478} (1997) L79.

\bibitem{adios} R.D. Blandford, M.C. Begelman {\it MNRAS} {\bf 303} (1999) L1.

\bibitem{mahadevan} R. Mahadevan {\it ApJ} {\bf 477} (1997) 585.

\bibitem{ggfr1fr2} G. Ghisellini, A. Celotti {\it A\&A} {\bf 379} (2001) L1.

\bibitem{ghi08}  G. Ghisellini et al. {\it MNRAS} {\bf 301} (1998) 451.

\bibitem{celotti}  A. Celotti, G. Ghisellini {\it MNRAS } {\bf 385} (2008) 283.

\bibitem{2155}  F. Aharonian et al. {\it ApJ Lett.}  {\bf 696} (2009) L150.

\bibitem{bllacs}  F. Tavecchio et al. {\it MNRAS}  {\bf 401} (2010) 1570 .

\bibitem{m87} F. Tavecchio, G. Ghisellini {\it MNRAS} {\bf 385} (2008) L98.

\bibitem{ghi05} G. Ghisellini et al. {\it A\&A} {\bf 432} (2005) 401.

\bibitem{0229} F. Tavecchio et al. {\it MNRAS} {\bf 399} (2009) L59.

\bibitem{katar} K. Katarzy{\'n}ski et al. {\it MNRAS} {\bf 368} (2006) L52.

\bibitem{abdobllacs} A. A. Abdo et al. {\it ApJ}  (2009) in press (arXiv:0910.4881).


\end{thebibliography}
\end{document}